\documentclass[11pt]{article}
\usepackage{amssymb} \usepackage{amsfonts} \usepackage{amsthm}
\usepackage{latexsym} \newtheorem{prop}{Proposition}
 \newtheorem{thm}{Theorem}
\newtheorem*{thm*}{Theorem} \newtheorem*{lemma*}{Lemma}
\newtheorem*{sv}{Stone-von Neumann Uniqueness Theorem}
 \theoremstyle{remark}
 
 \theoremstyle{definition}
 \newcommand{\abs}[1]{\mbox{$|#1|$}}

\newcommand{\norm}[1]{\mbox{$\| #1\|$}}
\newcommand{\hil}[1]{\mathcal{#1}} \newcommand{\alg}[1]{\mathcal{#1}}

\newcommand{\weyl}{\mathcal{A}[\mathbb{R}^{2}]}

 \title{Complementarity
  of representations in \\ quantum mechanics}
\author{Hans Halvorson \\
  {\small Department of Philosophy, Princeton University} \\
  {\small hhalvors@princeton.edu } } \date{}
\begin{document}
\maketitle 
\begin{abstract} 
  We show that Bohr's principle of complementarity between position
  and momentum descriptions can be formulated rigorously as a claim
  about the existence of representations of the CCRs.  In particular,
  in any representation where the position operator has eigenstates,
  there is no momentum operator, and vice versa.  Equivalently, if
  there are nonzero projections corresponding to sharp position
  values, all spectral projections of the momentum operator map onto
  the zero element.
\end{abstract}

\section{Introduction}
Niels Bohr's principle of complementarity has both a positive and a
negative tenet.  According to the positive tenet, a particle can have
\emph{either} a sharp position \emph{or} a sharp momentum.  According
to the negative tenet, a particle can never have \emph{both} a sharp
position \emph{and} a sharp momentum.  In this paper, I show that both
tenets of Bohr's complementarity principle correspond to
straightforward mathematical facts about the existence of
representations of the canonical commutation relations (CCRs).  In
particular, there is a (nonregular) ``position representation'' with a
complete set of position eigenstates, and there is a (nonregular)
``momentum representation'' with a complete set of momentum
eigenstates.  However, in any representation with propositions
ascribing a precise position value, there is no momentum operator, and
all propositions attributing a momentum value to the particle are
\emph{contradictory}.

As a foil to my account, I consider two criticisms --- one
corresponding to each tenet --- of Bohr's complementarity principle.
The criticism of the negative tenet is well-known: It is claimed that
Bohr makes a suspect inference from lack of joint measurability to
lack of joint reality.  The results of this paper, however, show that
one has no need to appeal to suspect philosophical doctrines in order
to provide a solid foundation for the negative tenet of
complementarity.  The criticism of the positive tenet --- although
less well-known --- is much more difficult to overcome.  In
particular, it is a simple mathematical fact that there are no states
in the standard Schr{\"o}dinger representation of the CCRs in which a
particle has a precise position or momentum.

However, this second criticism provides the key to unlocking both
sides of the complementarity principle: If we want to maintain that
particles can have precise positions or momenta, we must abandon one
of the assumptions that entails (via the Stone-von Neumann theorem)
the uniqueness of the Schr{\"o}dinger representation.  Once we drop
this assumption, we will see that there is a representation of the
CCRs which has the resources to describe precise position values, and
there is a representation of the CCRs which has the resources to
describe precise momentum values.  However, there is no single
representation of the CCRs that has the resources to describe
\emph{both} precise position values \emph{and} precise momentum
values.  (Along the way, I argue that the ``problem'' of inequivalent
representations is not peculiar to quantum field theory, but arises
already in elementary quantum mechanics.)

\bigskip \noindent {\it Preliminaries:} \smallskip \newline Let $\mu$
denote the Lebesgue measure on $\mathbb{R}$.  We say that a Borel
function $f$ from $\mathbb{R}$ to $\mathbb{C}$ is
\emph{square-integrable} just in case $\int
_{\mathbb{R}}\abs{f}^{2}d\mu <\infty$.  If $f,g$ are
square-integrable, we write $f\sim g$ just in case $\int
_{\mathbb{R}}\abs{f-g}^{2}d\mu=0$, and we let $[f]=\{ g: g\sim f \}$.
Let $L_{2}(\mathbb{R})$ denote the Hilbert space of equivalence
classes (mod $\sim$) of square-integrable functions from $\mathbb{R}$
into $\mathbb{C}$, with inner product given by $\langle [f],[g]\rangle
=\int _{\mathbb{R}}\overline{f}g\,d\mu$.  Let $\Sigma (\mathbb{R})$
denote the $\sigma$-algebra of Borel subsets of $\mathbb{R}$.  We
define an equivalence relation $\approx$ on $\Sigma (\mathbb{R})$ by
setting $S_{1}\approx S_{2}$ just in case $\mu (S_{1}-S_{2})+\mu
(S_{2}-S_{1})=0$.

Let $\iota$ be the identity function, $\iota (x)=x$, on $\mathbb{R}$.
Let $Q$ denote the self-adjoint operator defined on a dense domain of
$L_{2}(\mathbb{R})$ by $Q[f]=[\iota \cdot f]$, and let $P$ denote the
self-adjoint operator defined on a dense domain of $L_{2}(\mathbb{R})$
by $P[f]=-i[f']$.  (We set $\hbar =1$ throughout.)  Since $Q$ and $P$
satisfy the canonical commutation relation $[Q,P]=iI$, they can be
taken as corresponding, respectively, to the position and momentum
observables of a particle with one degree of freedom.  For $S\in
\Sigma (\mathbb{R})$, we define the projection operator $E^{Q}(S)$ on
$L_{2}(\mathbb{R})$ by setting $E^{Q}(S)[f]:=[\chi _{S}\cdot f]$,
where $\chi _{S}$ is the characteristic function of $S$.  In this
case, $E^{Q}$ gives the canonical spectral measure for the
self-adjoint operator $Q$.

Finally, let $\alg{L}$ denote the lattice of projection operators on
$L_{2}(\mathbb{R})$, equipped with the standard operations $\wedge
,\vee ,\perp$ corresponding respectively to subspace intersection,
span, and orthocomplement.

\section{A problem with position-momentum complementarity} \label{problem}
The complementarity between position and momentum is sometimes
mistakenly equated with the \emph{uncertainty} relation:
\begin{equation}
\left( \Delta _{\psi}Q \right) \cdot \left( \Delta _{\psi}P\right) \geq 1/2 \, ,
\label{heisenberg} \end{equation}
where $\Delta _{\psi}B$ is the dispersion of $B$ in $\psi$.  But
Eqn.~\ref{heisenberg} says nothing about when $Q$ and $P$ can
\emph{possess} values; at best, it only tells us that there is a
reciprocal relation between our \emph{knowledge} of the value of $Q$
and our \emph{knowledge} of the value of $P$.  To infer from this that
$Q$ and $P$ cannot simultaneously possess values would be to lapse
into positivism.
 
A more promising analysis of complementarity is suggested by Bub and
Clifton's \cite{bc} classification of ``no collapse'' interpretations
of quantum mechanics.  According to this analysis, we can think of
Bohr's complementarity interpretation as a no collapse interpretation
in which the measured observable $R$ and a state $\psi$ determines a
unique maximal sublattice $\alg{L}(\psi ,R)$ of the lattice $\alg{L}$
of all subspaces of the relevant Hilbert space.  $\alg{L}(\psi ,R)$
should be thought of as containing the propositions that have a
definite truth value when $R$ is measured in the state $\psi$.  In
particular, $\alg{L}(\psi ,R)$ always contains all propositions
ascribing a value to $R$ (i.e.~the spectral projections of $R$), and
$\psi$ can be decomposed into a mixture of 2-valued homomorphisms
(i.e.~``truth valuations'') on $\alg{L}(\psi ,R)$.  Thus, we can think
of $\psi$ as representing our \emph{ignorance} of the possessed value
of $R$.  We would then say that observables $R$ and $R'$ are
complementary just in case propositions attributing a value to $R'$
are never contained in $\alg{L}(\psi ,R)$ and vice versa.

But there is a serious difficulty in using this analysis to explicate
Bohr's notion of position-momentum complementarity.  In order to see
this, note that for any $S_{1},S_{2}\in \Sigma (\mathbb{R})$, if
$S_{1}\approx S_{2}$ then $(\chi _{S_{1}}\cdot f)\sim (\chi
_{S_{2}}\cdot f)$, and thus $E^{Q}(S_{1})[f]=E^{Q}(S_{2})[f]$.  Since
this is true for any $[f]\in L^{2}(\mathbb{R})$, it follows that
$E^{Q}(S_{1})=E^{Q}(S_{2})$ when $S_{1}\approx S_{2}$.  In particular
$E^{Q}(\{ \lambda \})=\mathbf{0}$, for any $\lambda \in \mathbb{R}$,
since $\{ \lambda \}\approx \emptyset$.  This enables us to formulate
a very simple ``proof'' that particles cannot have sharp positions.
\begin{description}
\item[{\it Mathematical Fact:}] $E^{Q}(\{ \lambda \})=\mathbf{0}$;
  i.e.~$E^{Q}(\{ \lambda \} )$ is the contradictory proposition.
\item[{\it Interpretive Assumption:}] $E^{Q}(\{ \lambda \} )$
  represents the proposition ``The particle is located at the
  point~$\lambda$.''
\item[{\it Conclusion:}] It is always false that the particle is
  located at $\lambda$.
\end{description}
What is more, it follows from the interpretive assumption that ``being
located in $S_{1}$'' is literally \emph{the same property as} ``being
located in $S_{2}$,'' whenever $S_{1}\approx S_{2}$.  Thus, any
attempt to attribute a position to the particle would force us to
revise the classical notion of location in space.

Halvorson~\cite{contin} argues that we can solve these difficulties by
reinterpreting elements of $\alg{L}$ as ``experimental propositions''
rather than as ``property ascriptions,'' and by introducing
non-countably additive (i.e.~non-vector) states on $\alg{L}$.  In
particular, suppose that we adopt the alternative interpretive
assumption:
\begin{description}
\item[{\it Interpretive Assumption-2:}] $E^{Q}(S)$ means ``A
  measurement of the position of the particle is certain to yield a
  value in $S$.''
\end{description}
Then, $E^{Q}(\{ \lambda \})=\mathbf{0}$ does \emph{not} entail that a
particle cannot be located at $\lambda$, but only that no position
measurement can be certain to show that it is located at $\lambda$.
Moreover, there is a non-countably additive state $h$ on $\alg{L}$
such that $h(E^{Q}(S))=1$ for \emph{all} open neighborhoods $S$ of
$\lambda$.  Thus, we could think of $h$ as representing a state in
which the particle is located at~$\lambda$.
  
However, this solution is not fully satisfactory.  In particular,
there is still no proposition in the ``object language'' $\alg{L}$
which can express the claim that the particle is located at $\lambda$.
[A pure state $h$ on $\alg{L}$ is countably additive if and only if
there is a unique minimal element $E\in \alg{L}$ such that $h(E)=1$.
Therefore, if we drop countable additivity, we allow for there to be
more states than can be described in the language $\alg{L}$ of the
theory.]  Thus, the standard language $\alg{L}$ of quantum mechanics
is incapable of describing particles as having precise positions or
momenta.  And, if we believe (as Bohr does) that particles \emph{can}
have precise positions or momenta, we are forced to conclude that
quantum mechanics is \emph{descriptively incomplete} (see
\cite{teller}).

\section{The Solution: Inequivalent Representations} \label{solution}
Why should we take $L_{2}(\mathbb{R})$ as the state space for a
particle with one degree of freedom?  And, why is the lattice
$\alg{L}$ of subspaces for $L_{2}(\mathbb{R})$ supposed to give us all
possible properties of a particle with one degree of freedom?  One
answer to these questions is that $L_{2}(\mathbb{R})$ and $\alg{L}$
supply the elements needed for an empirically adequate model of such a
particle.  However, this answer is not sufficient, because it leaves
open the possibility that there are inequivalent formalisms which
could also model the phenomena.  In order to supply a fully convincing
answer, we would need a uniqueness theorem which shows that \emph{any}
empirically adequate (or physically reasonable) model is equivalent to
the standard model.

Now, in the case of elementary quantum mechanics we do have a
uniqueness theorem.  In order to state this precisely, we first define
one parameter groups of unitary operators on $L_{2}(\mathbb{R})$ by
setting $U_{a}=\exp \{ iaQ\}$ and $V_{b}=\exp \{ ibP\}$ for all
$a,b\in \mathbb{R}$.  Then, as is well-known, these groups satisfy the
\emph{Weyl form} of the CCRs.
\begin{equation}
U_{a}V_{b}=e^{-iab}V_{b}U_{a}, \qquad \quad (a,b\in \mathbb{R}). \label{ccr}
\end{equation}
More generally, we say that any pair $(\{ U_{a}\} ,\{ V_{b}\} )$ of
one-parameter groups of unitary operators acting a Hilbert space
$\hil{H}$ give a representation of the Weyl form of the CCRs just in
case they satisfy Eqn.~\ref{ccr}.  Furthermore, we say that the
representation is \emph{irreducible} just in case no nontrivial
subspaces of $\hil{H}$ are left invariant by all operators $\{
U_{a},V_{b}:a,b\in \mathbb{R} \}$. We say that two representations
$(\{ U_{a}\} ,\{ V_{b}\} )$ and $(\{ \tilde{U}_{a} \} ,\{
\tilde{V}_{b}\} )$ on Hilbert spaces $\hil{H}$ and
$\tilde{\mathcal{H}}$ are \emph{unitarily equivalent} just in case
there is a unitary operator $W:\hil{H}\mapsto \tilde{\mathcal{H}}$
such that \begin{equation} WU_{a}W^{*}=\tilde{U}_{a}, \quad \qquad
  WV_{b}W^{*}=\tilde{V}_{b} ,\end{equation} for all $a,b\in
\mathbb{R}$.  Finally, a representation of the Weyl form of the CCRs
is said to be \emph{regular} just in case $a\mapsto U_{a}$ and
$b\mapsto V_{b}$ are continuous.  (Where $a\mapsto U_{a}$ is
continuous just in case $a\mapsto \langle \varphi ,U_{a}\varphi
\rangle$ is continuous for any $\varphi \in \hil{H}$.)  In this case,
it follows that there is a unique irreducible representation of the
CCRs.

\begin{sv}[\cite{vn}] Every irreducible regular representation of the 
  CCRs is unitarily equivalent to the Schr{\"o}dinger representation
  on $L_{2}(\mathbb{R})$.  \end{sv}

The Stone-von Neumann uniqueness theorem seems to force our
interpretive hand: There is a unique representation of the CCRs, and
the ``language'' $\alg{L}$ of that representation (i.e.~the lattice of
projection operators) either has no sentences that ascribe precise
positions or momenta to particles, or (if one thinks of projection
operators as ascribing properties) those sentences are necessarily
false.  Thus, if quantum mechanics is descriptively complete,
particles do not have precise positions or momenta.

However, this analysis is simply wrong.  The Stone-von Neumann theorem
cannot be used to prove that particles do not have precise positions
or momenta, because \emph{the regularity assumption begs the question
  against position-momentum complementarity}.  In particular, Stone's
theorem \cite[Thm.~5.6.36]{kr} entails that a representation of the
CCRs is regular if and only if the self-adjoint generators $Q$ of $\{
U_{a}:a\in \mathbb{R}\}$ and $P$ of $\{ V_{b}:b\in \mathbb{R}\}$ exist
on $\hil{H}$.  However, if complementarity is correct, then $Q$ and
$P$ cannot both have sharp values.  Why, then, do we need to assume
that both operators exist in one representation space?  What is more,
we saw in the previous section that if \emph{both} operators do exist,
then \emph{neither} can possess a sharp value.  Does this not give us
a reason to rethink the regularity assumption?

Let us formulate the previous argument more explicitly.  First, let DC
denote the claim that the language of quantum mechanics is
descriptively complete:
\begin{description} 
\item[{\it Descriptive Completeness}] (DC): A particle can have a
  property $E$ only if there is a corresponding projection operator
    $\hat{E}$ in some representation of the CCRs. \end{description} %
  Let SP denote the claim that particles \emph{can} have sharp
  positions, and let SM denote the claim that particles \emph{can}
  have sharp momenta.  Finally, let R denote the claim that any
  physically reasonable representation of the CCRs must be regular.
  Then the Stone-von Neumann theorem shows that:
\begin{quote}\begin{center} 
    DC $\wedge$ R $\Longrightarrow$ $\neg$SP $\wedge$ $\neg$SM
   \end{center}\end{quote} %
 However, this is equivalent to:
\begin{quote}\begin{center} DC $\wedge$ (SP $\vee$ SM) $\Longrightarrow$
    $\neg$R \end{center}\end{quote} Thus, it is logically coherent to
maintain the descriptive completeness of quantum mechanics along with
the claim that particles can have precise positions or momenta; to do
so, we must reject the assumption that any physically reasonable
representation of the CCRs needs to be regular.  [It should be noted
that the regularity assumption can be replaced with the assumption
that the Hilbert space $\hil{H}$ is separable (see \cite{summers}).
However, the warrant for separability is even shakier than the warrant
for regularity.]  In order to show that this logical possibility is
real, we now construct nonregular representations of the CCRs in which
there are \emph{contingent} statements attributing precise position
(or momentum) values to a particle.

\bigskip \noindent {\it 3.1 The Position Representation.}
\nopagebreak[3] \smallskip \newline \nopagebreak[3] Let
$l_{2}(\mathbb{R})$ denote the (nonseparable) Hilbert space of
square-summable functions from $\mathbb{R}$ into $\mathbb{C}$.  That
is, an element $f$ of $l_{2}(\mathbb{R})$ is supported on a countable
subset $S_{f}$ of $\mathbb{R}$ and $\norm{f}:=\sum _{x\in
  S_{f}}\abs{f(x)}^{2}<\infty$.  The inner product on
$l_{2}(\mathbb{R})$ is given by
\begin{equation} \langle f,g \rangle = \sum _{x \in S_{f}\cap
    S_{g}}\overline{f(x)}g(x) \, .
\end{equation} For each $\lambda \in \mathbb{R}$, let $\varphi
_{\lambda}$ denote the characteristic function of $\{ \lambda \}$.
Thus, the set $\{ \varphi _{\lambda}:\lambda \in \mathbb{R}\}$ is an
orthonormal basis for $l_{2}(\mathbb{R})$.  For each $a\in
\mathbb{R}$, define $U_{a}$ on the set $\{ \varphi _{\lambda}:\lambda
\in \mathbb{R}\}$ by
\begin{equation} U_{a}\varphi _{\lambda}:=e^{ia\lambda }\varphi
  _{\lambda } \, .\end{equation} Since $U_{a}$ maps $\{ \varphi
_{\lambda}:\lambda \in \mathbb{R}\}$ onto an orthonormal basis for
$l_{2}(\mathbb{R})$, $U_{a}$ extends uniquely to a unitary operator on
$\hil{H}$.  Similarly, define $V_{b}$ on $\{ \varphi
_{\lambda}:\lambda \in \mathbb{R} \}$ by
  \begin{equation}
V_{b}\varphi _{\lambda}:=\varphi _{\lambda -b} \, .\end{equation}
Then $V_{b}$ extends uniquely to a unitary operator on
$l_{2}(\mathbb{R})$.  Now, a straightforward calculation shows that, \begin{eqnarray}
U_{a}V_{b}\varphi
_{\lambda}&=&e^{-iab}V_{b}U_{a}\varphi _{\lambda} 
\, , \end{eqnarray}
for any $a,b\in \mathbb{R}$.  Thus, the operators $\{ U_{a}:a\in
\mathbb{R} \}$ and $\{ V_{b}:b\in \mathbb{R}\}$ give a
representation of the Weyl form of CCRs on $l_{2}(\mathbb{R})$.  

Furthermore, $\lim _{a\rightarrow 0} \langle \varphi _{\lambda},
U_{a}\varphi _{\lambda} \rangle =\lim _{a\rightarrow 0} e^{ia\lambda}
=1$, for any $\lambda \in \mathbb{R}$.  Thus, $a\mapsto U_{a}$ is
continuous, and Stone's theorem entails that there is a self-adjoint
operator $Q$ on $l_{2}(\mathbb{R})$ such that $U_{a}=\exp \{ iaQ\}$
for all $a\in \mathbb{R}$.  In particular,
\begin{equation} Q\varphi _{\lambda }=-i\lim _{a\rightarrow
    0}a^{-1}(U_{a}-I)\varphi _{\lambda} =-i\lim _{a\rightarrow
    0}a^{-1}(e^{ia\lambda} -1)\varphi _{\lambda}=\lambda \varphi
  _{\lambda} ,\end{equation} for each $\lambda \in \mathbb{R}$.  On
the other hand, we have:
\[ \langle \varphi _{\lambda} ,V_{b}\varphi _{\lambda} \rangle \:=\: 
\left\{ \begin{array}{ll} 0 & \mbox{when}\: b\neq 0, \\
    1 & \mbox{when}\:b=0 . 
\end{array} \right . \] 
Thus, $b\mapsto V_{b}$ is not continuous, and there is no self-adjoint
operator $P$ such that $V_{b}=\exp \{ ibP\}$ for all $b\in
\mathbb{R}$.  In other words, the momentum operator \emph{does not
  exist} in this representation.

\bigskip \noindent {\it 3.2 The Momentum Representation.}  \smallskip
\newline By means of a completely analogous construction, we can
obtain a representation of the CCRs in which, for each $\lambda \in
\mathbb{R}$, there is a vector $\varphi _{\lambda}$ such that
$P\varphi _{\lambda}=\lambda \varphi _{\lambda}$ (see
\cite{beaume,fannes}).  In this case, however, it is not possible to
define a position operator.

\bigskip Thus, we have shown that it \emph{is} possible to give a
standard Hilbert space description in which a particle has a precise
position in the continuum; and it \emph{is} possible to give a
standard Hilbert space description in which a particle has a precise,
numerical momentum value.  However, the representations we constructed
have a curious feature: In the position representation we cannot
define a momentum operator, and in the momentum representation we
cannot define a position operator. We now show that this
``complementarity'' between position and momentum holds in any
representation of the CCRs.

\begin{thm} In any representation of the Weyl form of the CCRs, if $Q$
  exists and has an eigenvector then $P$ does not exist.  If $P$
  exists and has an eigenvector then $Q$ does not exist.
  \label{disjoint} \end{thm}

\begin{proof}  We show that if there is a common
  eigenvector $\varphi$ for $\{ V_{b}:b\in \mathbb{R}\}$ then
  $a\mapsto U_{a}$ is not continuous.  (The other half of the theorem
  follows by symmetry.)  Indeed, if $\varphi$ is a common eigenvector
  for $\{ V_{b}:b\in \mathbb{R}\}$ then,
  \begin{equation} e^{iab}\langle \varphi ,U_{a}\varphi \rangle
    =\langle \varphi , V_{-b}U_{a}V_{b}\varphi \rangle =\langle
    \varphi ,U_{a}\varphi \rangle \, , \end{equation} for all 
$a,b\in \mathbb{R}$.  But this is possible only if 
$\langle \varphi ,U_{a}\varphi \rangle=0$ when
  $a\neq 0$.  Since $\langle \varphi ,U_{a}\varphi \rangle
  =1$ when $a=0$, it follows that $a\mapsto U_{a}$ is not continuous.   \end{proof}

Of course, this theorem also shows that the position and momentum
representations are inequivalent.  And it shows (using the
contrapositive of each part) that the momentum and position
representations are both inequivalent to the Schr{\"o}dinger
representation.  (This latter fact is already clear, though, since the
Hilbert spaces differ in dimension.)

\section{Nonexistent Quantities}
I have claimed that if there are eigenstates for the position
observable, then the momentum observable ``does not exist.''  What was
actually shown is that if there are eigenstates for the position
observable, then $b\mapsto V_{b}$ is not continuous; and so we cannot
reconstruct the momentum operator by taking the derivative
$-i(\partial V_{t}/\partial t)|_{t=0}$.  However, this failure of
continuity does not itself carry a natural physical interpretation.

In this section, I give another perspective on the complementarity
between position and momentum.  In particular, we shall see that, in
\emph{every} representation of the CCRs, there are projection-valued
measures $E^{Q},E^{P}$ representing the position and momentum
observables.  However, the properties (in particular, the sets of
measure zero) of these measures vary between representations.  Thus,
the complementarity between position and momentum can be formulated as
a claim about the relation between $E^{Q},E^{P}$ in representations of
the CCRs.  In particular, our main result shows that if $E^{Q}(\{
\lambda \} )\neq \mathbf{0}$ for some $\lambda \in \mathbb{R}$ then
$E^{P}(\mathbb{R})=\mathbf{0}$.  That is, if it is possible for a
particle to have a point location, then it is impossible for that
particle to have any momentum value in $\mathbb{R}$.

\bigskip \noindent {\it 4.1 Preliminaries.} \smallskip \newline %
Let $BC(\mathbb{R})$ denote
the $C^{*}$-algebra of continuous functions from $\mathbb{R}$ into
$\mathbb{C}$ that are bounded in the norm
\begin{equation}
\norm{f}=\sup _{x\in \mathbb{R}}\abs{f(x)} .  \end{equation}
Elements of $BC(\mathbb{R})$ are called \emph{bounded continuous
  functions}.  Recall that the space of pure states of $BC(\mathbb{R})$ (equipped
with the weak* topology) is homeomorphic to the Stone-\v{C}ech
compactification of $\mathbb{R}$.  In particular, for each pure state
$\omega$ of $BC(\mathbb{R})$ there is a unique ultrafilter $\alg{U}$
on $\mathbb{R}$ such that $\omega (f) =\lim _{\mathcal{U}} f$.  

Let $X$ be a Hausdorff topological space, and let $\pi$ be a mapping
of $\mathbb{R}$ into $X$.  We say that $(\pi ,X)$ is a
\emph{compactification} of $\mathbb{R}$ just in case $X$ is compact,
and $\pi$ is a continuous embedding of $\mathbb{R}$ onto a dense
subset of $X$.  [We do not require $\pi$ to be a homeomorphism of
$\mathbb{R}$ onto $\pi (\mathbb{R})$.]  There is a one-to-one
correspondence between $C^{*}$-subalgebras of $BC(\mathbb{R})$ and
compactifications of $\mathbb{R}$ (see \cite[p.~16]{gamelin}).  In
particular, since each $\lambda \in \mathbb{R}$ gives rise to a
principal ultrafilter $\alg{U}_{\lambda}\,$ on $\mathbb{R}$, there is
a natural injection of $\mathbb{R}$ into the space $\sigma (\alg{A})$
of pure states of $\mathcal{A}$.  Furthermore, $\sigma (\alg{A})$ is
compact, and $\alg{A}$ is naturally isomorphic to the continuous
functions $C(\sigma (\alg{A}))$ on $\sigma (\alg{A})$.  Thus, a
subalgebra of $BC(\mathbb{R})$ is naturally isomorphic to the
continuous functions on a compact space $X=\mathbb{R}\cup (X\backslash
\mathbb{R})$, where the elements in $X\backslash \mathbb{R}$ can be
thought of as ``points at infinity.''

We now define a specific $C^{*}$-subalgebra of $BC(\mathbb{R})$.  For
each $a\in \mathbb{R}$, let $u_{a}$ denote the function given by
$u_{a}(x)=e^{iax},\; (x\in \mathbb{R})$.  Let $AP(\mathbb{R})$ denote
the $C^{*}$-subalgebra of $BC(\mathbb{R})$ generated by $\{ u_{a}:a\in
\mathbb{R} \}$.  That is, $AP(\mathbb{R})$ consists of uniform limits
of trigonometric polynomials of the form
\begin{equation}
f(x)=\sum _{j=1}^{n}c_{j}e^{ia_{j}x} ,\end{equation}
where $c_{j}\in \mathbb{C}$ and $a_{j}\in \mathbb{R}$.  Elements of
$AP(\mathbb{R})$ are called \emph{almost periodic functions}.   The 
compactification $b\mathbb{R}$ of $\mathbb{R}$ corresponding to
$AP(\mathbb{R})$ is called the \emph{Bohr compactification} of
$\mathbb{R}$ (after the mathematician Harald Bohr). Thus,
$AP(\mathbb{R})$ is isomorphic to $C(b\mathbb{R})$.  

There is a binary operation $\hat{+}$ on $b\mathbb{R}$ such that
$(b\mathbb{R},\hat{+},0)$ is a compact topological group in which
$(\mathbb{R},+,0)$ is embedded as a dense subgroup
\cite[p.~30]{rudin1}.  Let $\mu$ denote the unique (normalized)
translation-invariant measure (i.e.~\emph{Haar measure}) on the Borel
$\sigma$-algebra $\Sigma (b\mathbb{R})$ of $b\mathbb{R}$.  Now define
the \emph{invariant mean} $\omega _{\mu}$ on $C(b\mathbb{R})$ by setting
\begin{equation} \omega _{\mu}(f):=\int _{b\mathbb{R}}f \, d\mu \, ,
  \quad \qquad (f\in C(b\mathbb{R})).\end{equation}  
The invariant mean can also be defined explicitly by
\begin{equation}
\omega _{\mu}(f)=\lim_{N\rightarrow \infty}\frac{1}{2N}\int
_{-N}^{N}f(x)\,dx\, ,\quad \qquad (f\in AP(\mathbb{R})). \end{equation}
Clearly, $\omega _{\mu}(u_{a})=0$ for
all $a\neq 0$; and since $\{ u_{a}:a\in \mathbb{R}\}$ is linearly dense
in $C(b\mathbb{R})$, it follows that $\omega _{\mu}$ is the unique
state of $C(b\mathbb{R})$ with this property.

\bigskip \noindent {\it 4.2 The Weyl algebra} \smallskip \newline %
There is a unique minimal $C^{*}$-algebra $\weyl$ containing two
one-parameter groups $\{ U_{a}:a\in \mathbb{R}\}$ and $\{ V_{b}:b\in
\mathbb{R}\}$ of unitary operators obeying the Weyl form of the CCRs
\cite{mstv}.  The abelian $C^{*}$-subalgebra of $\weyl$ generated by
$\{ U_{a}:a\in \mathbb{R}\}$ is naturally isomorphic to
$C(b\mathbb{R})$ $(\approx AP(\mathbb{R}))$, and the same is true of
the $C^{*}$-subalgebra generated by $\{ V_{b}:b\in \mathbb{R}\}$.

Recall that a representation $(\pi ,\hil{H})$ of a $C^{*}$-algebra
$\alg{A}$ consists of a $*$-homo{-}morphism $\pi$ of $\alg{A}$ into
the $C^{*}$-algebra of bounded linear operators on $\hil{H}$.  Thus,
the irreducible representations of the Weyl form of the CCRs are in
one-to-one correspondence with the irreducible representations of
$\weyl$.  Recall also that two representations $(\pi ,\hil{H})$ and
$(\tilde{\pi} ,\tilde{\mathcal{H}})$ of $\alg{A}$ are \emph{unitarily
  equivalent} just in case there is a unitary operator
$W:\hil{H}\mapsto \tilde{\mathcal{H}}$ such that $W\pi
(A)W^{*}=\tilde{\pi}(A)$ for all $A\in \alg{A}$.

\begin{prop} For any irreducible representation $(\pi ,\hil{H})$ of
  $\alg{A}[\mathbb{R}^{2}]$ there is a corresponding pair
  $(E^{Q},E^{P})$ of projection-valued measures on $b\mathbb{R}$ such
  that
\begin{equation} \pi (U_{a})=\int
      _{b\mathbb{R}} u_{a}(\lambda )\,dE^{Q}_{\lambda} , \qquad \quad \pi
    (V_{b})=\int _{b\mathbb{R}}u_{b}(\lambda )\,dE^{P}_{\lambda}
    , \label{integrate} \end{equation}
for all $a,b\in \mathbb{R}$.  
\label{pv}  \end{prop}

\begin{proof}  Define a pair $\alpha ^{Q},\alpha ^{P}$ of representations of
  $C(b\mathbb{R})$ on $\hil{H}$ by setting $\alpha ^{Q}(u_{a})=\pi
  (U_{a})$ and $\alpha ^{P}(u_{a})=\pi (V_{a})$ for all $a\in
  \mathbb{R}$.  The result then follows immediately from Theorem 5.2.6
  of \cite{kr}.  \end{proof}

The heuristic integrals in Eqn.~\ref{integrate} can be interpreted
rigorously as quantified statements about families of complex Borel
measures on $b\mathbb{R}$.  In particular, for any vector $\varphi \in
\hil{H}$ we define a measure $\mu _{\varphi}$ on $b\mathbb{R}$ by
setting
\begin{equation}
\mu _{\varphi}(S):= \langle \varphi ,E^{Q}(S)\varphi \rangle ,\quad
\qquad (S\in \Sigma (b\mathbb{R})) .\end{equation}
Then, we define the left part of Eqn.~\ref{integrate} as the
statement:  
\begin{equation}
\langle \varphi ,\pi (U_{a})\varphi \rangle = \int
_{b\mathbb{R}}u_{a}(\lambda )\, d\mu _{\varphi}(\lambda ), \qquad
\forall \varphi \in \hil{H} . \end{equation}
The right part of Eqn.~\ref{integrate} makes a similar statement about
complex Borel measures defined in terms of $E^{P}$ and vectors in $\hil{H}$.

We also claim (without proof) that $a\mapsto \pi (U_{a})$ is
continuous iff. $E^{Q}(\mathbb{R})=\mathbf{I}$.  In this case, $\{
E^{Q}_{\lambda}:\lambda \in \mathbb{R}\}$ gives the spectral
resolution of $Q\,$:
 \begin{equation} Q=\int _{\mathbb{R}} \lambda
  \;dE^{Q}_{\lambda} .
\end{equation} Similarly, $b\mapsto \pi (V_{b})$ is continuous if and
only if $E^{P}(\mathbb{R})=\mathbf{I}$, in which case $\{
E^{P}_{\lambda}:\lambda \in \mathbb{R}\}$ gives the spectral
resolution of $P$.

Irreducible representations of $\weyl$ are in one-to-one
correspondence with pure states of $\weyl$.  In particular, for any
pure state $\omega$ of $\weyl$, the GNS construction provides an
irreducible representation $(\pi _{\omega},\hil{H}_{\omega})$ of
$\weyl$ such that $\omega$ is represented by a cyclic vector $\Omega
\in \hil{H}_{\omega}$ \cite[Thm.~4.5.2; Thm.~10.2.3]{kr}.  In fact, we
can directly produce both the position and momentum representations by
means of the GNS construction: For each $\lambda \in \mathbb{R}$,
there is a unique pure state $\omega _{\lambda}$ of $\weyl$ such that
$\omega _{\lambda} (U_{a})=e^{ia\lambda}$ for all $a\in \mathbb{R}$;
moreover, it follows that $\omega _{\lambda}(V_{b})=0$ when $b\neq 0$
(see \cite{beaume}).  Thus, $\omega _{\lambda}(A)=\langle \varphi
_{\lambda},A\varphi _{\lambda}\rangle$ whenever $A=U_{a}$ or
$A=V_{a}$, for some $a\in \mathbb{R}$.  Since $\{ V_{b}\varphi
_{\lambda}:b\in \mathbb{R}\}$ is also linearly dense in
$l_{2}(\mathbb{R})$, the position representation is unitarily
equivalent to the GNS representation induced by $\omega _{\lambda}$
\cite[Prop.~4.5.3]{kr}.  A completely analogous construction can be
used to obtain a GNS representation that is unitarily equivalent to
the momentum representation.
  
\begin{thm} Let $(\pi ,\hil{H})$ be an irreducible representation of
  $\alg{A}[\mathbb{R}^{2}]$, and let $(E^{Q},E^{P})$ be the
  corresponding pair of projection-valued measures.  If $E^{Q}(\{
  \lambda \} )\neq \mathbf{0}$ for some $\lambda \in \mathbb{R}$ then
  $E^{P}(\mathbb{R})=\mathbf{0}$.  If $E^{P}(\{ \lambda \} )\neq
  \mathbf{0}$ for some $\lambda \in \mathbb{R}$ then
  $E^{Q}(\mathbb{R})=\mathbf{0}$. \end{thm}

\begin{proof} The proof splits into two parts: (1.) If $E^{Q}(\{
  \lambda \} )\neq \mathbf{0}$ then $(\pi ,\hil{H})$ is unitarily
  equivalent to the position representation.  (2.) If $(\pi ,\hil{H})$
  is equivalent to the position representation then
  $E^{P}(\mathbb{R})=\mathbf{0}$.
  
  (1.) Let $(\pi ,\hil{H})$ be an irreducible representation of
  $\weyl$ such that $E^{Q}(\{ \lambda \} )\neq \mathbf{0}$. Let
  $\varphi$ be a unit vector in the range of $E^{Q}(\{ \lambda \} )$,
  and define a Borel measure $\nu$ on $b\mathbb{R}$ by setting
  \begin{equation}
\nu (S):=\langle \varphi ,E^{Q}(S)\varphi \rangle ,\quad \qquad (S\in
\Sigma (b\mathbb{R})). \end{equation}
Then $\nu$ is concentrated on $\{ \lambda \}$, and Prop.~\ref{pv} 
entails that \begin{equation}
\langle \varphi ,\pi (U_{a})\varphi \rangle  = \int _{b\mathbb{R}}
u_{a} \,d\nu =u_{a}(\lambda )=e^{ia\lambda} .\end{equation}
Since $(\pi ,\hil{H})$ is irreducible, $\varphi$ is cyclic for $\pi$.
It follows then from the uniqueness of the GNS
  representation \cite[Prop.~4.5.4]{kr}, that $(\pi ,\hil{H})$ is
  unitarily equivalent to the GNS representation of $\weyl$ induced by the state
  $\omega _{\lambda}$, and thus to the position representation.
  
  (2.) Suppose that $(\pi ,\hil{H})$ is unitarily equivalent to the
  position representation.  In particular, we can suppose that
  $\hil{H}$ $(=l_{2}(\mathbb{R}))$ is spanned by an orthonormal family
  of vectors $\{ \varphi _{\lambda}:\lambda \in \mathbb{R} \}$ where
  $\pi (V_{b})\varphi _{\lambda}=\varphi _{\lambda -b}$ for all
  $b,\lambda \in \mathbb{R}$.  Thus, to show that
  $E^{P}(\mathbb{R})=\mathbf{0}$, it will suffice to show that
  $E^{P}(\mathbb{R})\varphi _{\lambda}=0$ for all $\lambda \in
  \mathbb{R}$.  Fix $\lambda$ and let $\varphi = \varphi _{\lambda}$.
  Define a Borel probability measure $\nu$ on $b\mathbb{R}$ by setting
\begin{equation}
\nu (S) :=\norm{ E^{P}(S)\varphi }^{2} =\langle \varphi 
,E^{P}(S)\varphi \rangle ,\qquad \quad (S\in
\Sigma (b\mathbb{R})) .\end{equation}
Define a state $\omega _{\nu}$ of $C(b\mathbb{R})$ by
\begin{equation}
\omega _{\nu}(f) := \int _{b\mathbb{R}}f\, d\nu ,\quad \qquad (f\in
C(b\mathbb{R})). \end{equation}
Then Prop.~\ref{pv} entails that \begin{equation} \omega _{\nu}(u_{a})
  = \int _{b\mathbb{R}}u_{a} \,d\nu = \langle \varphi ,\pi
  (V_{a})\varphi \rangle =0 ,\end{equation} for all $a\in \mathbb{R}$.
Since $\{ u_{a}:a\in \mathbb{R}\}$ generates $C(b\mathbb{R})$, it follows that
$\omega _{\nu}$ is the invariant mean.  By the Riesz representation
theorem \cite[Thm.~2.14]{rudin}, there is a unique probability measure
on $b\mathbb{R}$ corresponding to each state of $C(b\mathbb{R})$.  Thus, $\nu$ is the Haar
measure on $b\mathbb{R}$, and it follows from translation-invariance (and countable-additivity) 
that $\nu (\mathbb{R})=0$.
\end{proof}

\section{Conclusion}

It is well-known that the existence of inequivalent representations
raises significant issues for the interpretation of quantum
\emph{field} theory.  However, it would be wrong to think (on the
basis of the Stone-von Neumann uniqueness theorem) that the issue of
inequivalent representations has no significance for elementary
quantum mechanics.  Indeed, it is only by employing nonregular
representations of the CCRs that we can make sense of Bohr's views
about position-momentum complementarity.

Still, one might wonder whether nonregular representations of the CCRs
have any \emph{empirical} relevance.  In particular, don't we have
purely empirical grounds for preferring the Schr{\"o}dinger
representation to these representations?  But this question betrays a
misunderstanding of the nature of representations.  The abstract
algebra of observables $\weyl$ carries the full empirical content of
the quantum theory of a single particle (with one degree of freedom).
(We have not mentioned dynamics, but this can also be defined in a
representation-indendent manner.)  In particular, $\weyl$ has enough
observables to describe our measurement procedures, and enough states
to describe each laboratory preparation.  Thus, a representation does
not add anything further to this already given empirical content, and
one representation cannot be preferable to another on empirical
grounds alone.

However, by saying that any two representations of $\weyl$ are
empirically equivalent, we are not committing ourselves to the further
claim that there are no \emph{physically significant} differences
between representations.  In particular, each representation of
$\weyl$ comes with a certain set of ``sharp propositions''
(i.e.~projection operators) which can be used to describe ``how things
are'' independent of our experimental interventions.  Moreover,
different representations give very different stories about how things
are.  For example, while the Schr{\"o}dinger representation says that
two locations are the same if they differ only by a set of Lebesgue
measure zero, the position representation permits us to maintain the
classical picture of the spatial continuum in which particles can have
precise positions.

\bigskip {\it Acknowledgments:} Thanks to Rob Clifton for many
discussions and helpful comments.  I am also indebted to John Earman
and Laura Ruetsche for their insightful analyses of the issues raised
by the existence of inequivalent representations.

\end{document}